# MIS on the fly
(Extended Abstract)

Yehuda Afek [*]    Noga Alon [†]    Ziv Bar-Joseph [‡]

**Abstract**

Humans are very good at optimizing solutions for specific problems. Biological processes, on the other hand, have evolved to handle multiple constrained distributed environments and so they are robust and adaptable. Inspired by observations made in a biological system we have recently presented a simple new randomized distributed MIS algorithm [1]. Here we extend these results by removing a number of strong assumptions that we made, making the algorithms more practical. Specifically we present an $O(\log^2 n)$ rounds synchronous randomized MIS algorithm which uses only 1 bit unary messages (a beeping signal with collision detection), allows for asynchronous wake up, does not assume any knowledge of the network topology, and assumes only a loose bound on the network size. We also present an extension with no collision detection in which the round complexity increases to $(\log^3 n)$. Finally, we show that our algorithm is optimal under some restriction, by presenting a tight lower bound of $\Omega(\log^2 n)$ on the number of rounds required to construct a MIS for a restricted model.

**Intended for Regular Presentation**.
**Contact Author:**
Ziv Bar-Joseph
Email:        zivbj@cs.cmu.edu

[*]The Blavatnik School of Computer Science, Tel-Aviv University, Israel 69978. afek@post.tau.ac.il

[†]School of Mathematics, & The Blavatnik School of Computer Science, Tel-Aviv University, Israel 69978. nogaa@post.tau.ac.il

[‡]School of Computer Science, Carnegie Mellon University, Pittsburgh, PA 15213, USA. zivbj@cs.cmu.edu

# 1 Introduction

Biological processes are often distributed both at the molecular level (within cells) and at the cellular level (between cells). Regulatory and signaling networks, which operate within cells, utilize several independent proteins (termed receptors) to sense the environment. These proteins coordinate programs that lead to the expression of new proteins which distributively handle such processes as DNA replication [19] and stress response [6]. Communication between cells is used to fight invading pathogens [10], reach decisions regarding cell fate [3] and form new blood vessels. We and others have recently shown that distributed computing principles including synchronization, backup and fault tolerance are commonly used by these biological processes [9, 8, 11]. An interesting question that arises is whether there are any insights that we can gain from studying these biological processes that can help us improve distributed computing algorithms.

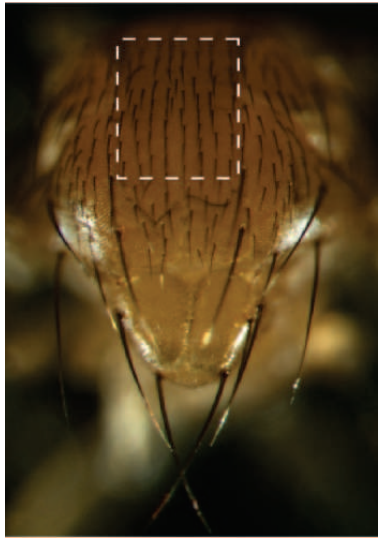

Figure 1: SOP selection in flies. The tiny bristles on the fly's forehead (inside the dashed rectangle) are each connected to a neuron. The spacing between them is achieved by an MIS selection like process in which a subset of the cells is selected and these inhibit all their neighbors.

In a recent paper [1] we have answered this question in the affirmative. We studied a specific biological process, sensory organ precursors (SOP) selection, which takes place during the development of the fly's brain. SOPs are cells that later become small bristles on the fly's forehead (Figure 1) and are used for sensing. It has been known for decades that there is a very strict spacing pattern required for the placement of these SOP cells. Specifically, in a highly accurate process cells that decide to become SOPs inhibit their neighboring cells so that each SOP is surrounded by non SOP cells. The process is stochastic, initially all cells have the potential to become SOP cells. The problem of selecting SOPs is very similar to the problem of distributively selecting a Maximal Independent Set (MIS).

An MIS is a *maximal* set of nodes in a network such that no two of them are neighbors. Since the set is maximal every node in the network is either in the MIS or a neighbor of a node in the MIS. The problem of distributively selecting an MIS has been extensively studied in various models [2, 4, 22, 14, 15, 16, 17, 21, 20, 27] and has many applications in networking, and in particular in radio sensor networks. Some of the practical applications include the construction of a backbone for wireless networks, a foundation for routing and for clustering of nodes, and generating spanning trees to reduce communication costs [22, 27].

Many probabilistic algorithms have been suggested in the past three decades to select a MIS in a network, however all either assume nodes know the network topology to a certain extent, and/or that communication between nodes is by messages of logarithmic length in the network size. In contrast, as we have shown, flies solve the MIS problem without knowing the network topology nor sending multi bit messages. Inspired by a detailed study of the fly selection process we derived [1] a $O(\log^2 n)$ rounds synchronous MIS algorithm which does not require any knowledge about the topology except for a rough upper bound on the total number of nodes and uses only one bit unary messages (just a signal, similar to a beep as in [5]) with an overall optimal message complexity.

While biological systems are not always optimal when it comes to run time complexity, they are usually very robust and adaptable. These systems usually use simple basic building blocks for the communication between cells or molecules. Software systems, on the other hand, are often sensitive to modeling assumptions and can fail when these assumptions do not hold. It is thus of interest to study the basic biological communication model in more detail and to derive algorithms that, even if they do not mimic directly the activity of the biological systems, can accommodate the limited set of assumptions they rely on. In this paper we introduce *the fly* model in which we formally capture our assumptions regrading the a biological communication model. Based on this model we extend our algorithm from [1] in several ways making it applicable to a wider set of applications. First, in Section 4, we remove its assumption that all nodes wake up simultaneously at the same round, while maintaining its complexities. Then, in Section 5, we extend it to the beeping model [5], where unlike our original model nodes cannot listen while broadcasting a signal. That is, our original algorithm assumes a beeping with collision detection model and we extend it to remove the collision detection assumption, paying a factor of $\log n$ in its round complexity. In a third variation on our algorithm we loosen its assumption about the knowledge of $n$ the number of nodes in the network. Our original algorithm assumes knowledge of a polynomial upper bound (in the actual number) on the number of nodes. Our new variant (Section 6) allows our upper bound, $N$, to be exponential in the actual number $n$ and maintaining a $O(\log n \log N)$ rounds time complexity. Finally, in Section 7 we prove a lower bound of $\Omega(\log^2 n)$ on the number of rounds, under the restriction (which is obeyed by all the algorithms we present), that all the nodes broadcast with the same probability in each round (but possibly different from round to round).

## 2 Related work

While several methods were suggested for solving MIS in the case of symmetric processors, these methods have mostly relied on nodes knowing the set of active neighbors each has at each stage of the execution. Moscibroda and Wattenhofer [21] were the first to consider the more realistic setting in which such information is not known. Similar to our model they consider MIS selection in multi-hop networks where only an upper bound is known regarding the number of nodes in the network and with asynchronous wake up. Unlike our model they do not assume that nodes are equipped with collision detectors. However, their model allows for (and their algorithm uses) messages whose size is a function of the number of nodes in the network. In contrast, for the unary message model, collision detectors maybe more realistic [23]. For that model Moscibroda and Wattenhofer present a $(\log^2 n)$ algorithm though, unlike our algorithm their algorithm is restricted to a specific type of graph (unit disk graph). In [24] Schneider and Wattenhofer presented a $O(\log^* n)$ algorithm for the more general (though still restricted) polynomially bounded growth graphs. However, that algorithm does not assume that collisions occur and so, as the authors write, is less appropriate for wireless networks.

In [20] the importance of the bit complexity in distributed computing algorithm is discussed. They present a $O(\log n)$ algorithm for MIS using only constant size messages for a total of $O(\log n)$ bits per channel. However, unlike our model they do not account for collisions and assume that nodes can count the number of neighbors they have and distinguish between messages sent by different neighbors. In addition, their algorithms uses messages that are larger than 1 bit and while each channel only delivers $O(\log n)$ bits, the overall message complexity may be much larger than $O(n)$ depending on the connectivity of the network. In contrast, our algorithm has an overall linear message complexity.

Several papers discuss reasonable assumptions regarding collision detection and wake-up models in wireless sensor networks. These were recently classified based on the strength of the assumptions involved [12]. It was argued that collisions can be affected even by nodes that are not necessarily in direct contact with the receiving node [23]. However, these models refer to the ability to receive complete packets. If the only requirement is to be able to sense the medium to determine whether a message was sent or not, both the range of each node and its ability to identify collisions is increased. Note that in the fly model, any message that is sent contains the same information. This type of communication was recently termed the beeping model [5] and in that paper the authors presented an algorithm for interval coloring in this model assuming no collision detection. Their algorithm runs in $O(\Delta \log n)$ rounds where $\Delta$ is the maximal degree of any node in the network. In contrast, we present an $O(\log^3 n)$ rounds algorithm for MIS in this model for an arbitrary degree graph.

## 3 Model

We assume a synchronous communication model on an arbitrary graph $G = (V, E)$ where the vertices $V$ represent processors and the edges represent pairs of processors that hear each other. This is the same model as the discrete model in [5], except for collision detection. The model assumed in this paper, called *the fly* model, applies to both, our understanding of how cells interact as well as to radio sensor networks. Specifically, we assume the following:

1. Synchronous rounds - While in biology events are not fully synchronized, communication is via physical (cell-cell or protein-protein) interaction and so there exists an upper bound on message delay. Consequently we assume a synchronous network, as the discrete model in [5]. Nodes are asynchronous except that the transmission slots are discrete and start at the same time in all the nodes. It is easy to transform the results also to the un-slotted case, where all slots are of the same length but their starting times are not synchronized, the round complexity of the un-slotted version is twice (factor of 2) that of the corresponding slotted version.

2. Asynchronous wake-up: Several processes are initiated by cells when they receive signals from neighboring cells or from the environment. Thus the algorithm is initiated by an arbitrary subset of the nodes that wake up at arbitrary rounds, obviously not necessarily all together. Notice, that under this assumption the time complexity of the algorithm is at least the network diameter, the number of rounds it would take a woken up node to wake up all other nodes. However, for our time complexity we consider for each node the number of rounds it has been active in the algorithm. We discuss this in more details below.

3. No knowledge about the network topology or the number of neighbors: Cells physically interact, however, cells cannot tell exactly how many neighbors they have since the geometry of the interaction depends on the cell shape which can change over time (though a bound on this number is easy to derive). Correspondingly we assume a network in which nodes do not know the number of neighbors each has, nor do they have any other knowledge about the network topology. Except, nodes are given a polynomial upper bound on the actual number of nodes in the network, and in Section 6 we weaken this assumption and nodes know only some possibly exponential bound on the actual number of nodes in the network.

4. Beeping communication: Cells communicate by secreting certain proteins that are sensed ("heard") by neighboring cells [4]. This is similar to a node in a radio network transmitting a carrier signal which is sensed ("heard") by its neighbors. Thus we assume nodes communicate only by transmitting such signals, called beeps. All nodes within range of a beep transmitting station hear the beep if they listen.

5. Collision detection: Cells can distinguish silence from the case that at least one neighbor is transmitting a beep. Depending on the specific process cells can either send and receive beeps at the same round or cannot receive at the time that they transmit a beep, hence can only hear a beep if they are not transmitting at the same time [3]. In the corresponding network model we assume that in each round a node may broadcast a beep-signal or stay silent, and in each round a node hears if any of its neighbors (not including itself) has broadcasted or not. A node hearing a beeping signal cannot tell which of its neighbors beeped, or if one or more have beeped. We distinguish between two variations: whether a node can listen at the same round in which it is beeping or not as in [5].

We define the *time (round) complexity* in our model as follows: consider for each node the active time as the number of rounds it has been active in the algorithm. I.e., from the first round in which the node woke up (spontaneously or by a neighbor) until it and its immediate neighbors exit the algorithm. The time complexity is then the maximum active time (expected) over all the nodes. The *message complexity* is the expected total number times nodes broadcast a signal/beep to their neighbors.

This model is appropriate for many distributed computing applications, most notably ad-hoc sensor networks which operate in topologies that are initially unknown, are required to conserve energy and can more accurately detect (beeping) signals when compared to larger messages [23].

## 4 MIS algorithm for asynchronous wake-up

We start by presenting an algorithm similar to the one presented in [1], for the beeping with collision detection model. While in [1] we assumed that all nodes wake up simultaneously at the same round, here we allow for asynchronous wake-up. In addition, we provide in this section a new message complexity analysis, showing it is optimal. In [1] we considered the number of messages sent between nodes, i.e., a beep broadcast by a node with $d$ neighbors that are still active in the algorithm contributes $d$ to the message complexity. Here we count each local broadcast of a beep as 1 regardless of how many nodes hear this beep.

Intuitively, if we were told that the graph is nearly a clique, i.e., the degree of almost all the nodes is $n$, then we would let each node try to be a MIS member with probability $1/n$, and we would repeat this trial $C \log n$ times for some constant $C$. If in any trial exactly one node selected itself, then it becomes a member of the MIS and would deactivate all its neighbors. On the other hand if we were told that the degree of each node is 2 (network is a collection of rings), then we would let each node attempt to become a member in the MIS with probability $1/2$, and again we would repeat this $C \log n$ times. Roughly, since we have no knowledge about node degrees, our algorithm sweeps the space of possible node degrees from the high degree nodes (trials with small probabilities) to the lower degree nodes.

The algorithm is given in Figure 2. The algorithm proceeds in $\log n$ phases ($\log \Delta$ if an upper bound $\Delta$ on the largest node degree is given), each consisting of $M \log n$ steps where $M$ is an appropriately chosen absolute constant $= 34$ as given in [1]. Assuming all nodes start simultaneously at the same round, each step in phase $i$ consists of two exchanges. In the first exchange nodes beep with probability $1/2^{\log n - i} = 2^i/n$, and in the second exchange a node that beeped in the first and did not hear a beep from any of its neighbors, beeps again, telling its neighbors it has joined the MIS and they should become inactive and exit the algorithm.

To remove the assumption that all nodes start together nodes that woke up spontaneously propagate a wave of wake-up beep throughout the network. Upon hearing the first beep, which must be the wake up beep, a node broadcasts the wake up beep on the next round, and then waits one round to ensure none of its neighbors is still waking up. Each exchange of the original algorithm now consists of 3 rounds. Nodes listen in all three rounds to incoming messages. During the second round of the first exchange each active node broadcasts a message to its neighbors with probability $p_i$ where $p_i$ increases with $i$, as discussed above (so that the probability is initially very low and it increases with additional iterations). The second exchange also takes three rounds. A node that has broadcasted a message in the first exchange joins the MIS if none of its neighbors had broadcasted in any of the three round of the first exchange. Such node broadcasts again a message in the second round of the second exchange telling its neighbors to terminate the algorithm. Clearly replacing each exchange with 3 rounds does not affect the asymptotic round complexity of the algorithm. However, we need to show that it does not affect the safety lemmas and the fact that when the algorithms terminates all processors are either in the MIS or connected to a node in the MIS with high probability. We will start by proving that the termination lemma from [1], which relies on the fact that all neighbors are using the same probability distribution in each exchange, still holds.

---

/* $n$ is a given upper bound on the actual number of nodes in the network */

    Wait to hear the first wake up beep
    **If** received a beep (the first)
        **then** broadcast wake up beep to all neighbors    /* Make sure your neighbors wake up too
    Wait one round                                     /* While your neighbor(s) wake up their neighbors
    **for** $i = 0$ to $\log n$ **do:**                                   /* $\log n$ phases
      **for** $j = 0$ to $M \log n$ **do:**                             /* $M \log n$ steps
        **\*\* exchange 1 \*\* with 3 rounds**
      Listen for 1 round                                                      /* round 1
      $v = 0$
      With probability $\frac{1}{2^{\log n - i}}$ broadcast signal and set $v = 1$ ;           /* round 2
      Listen for 1 round                                                      /* round 3
      **If** heard a signal in any round of exchange 1 **then** $v = 0$
        **\*\* exchange 2 \*\* with 3 rounds**
      Listen for 1 round                                                      /* round 1
      **If** $v == 1$ **then** Broadcast signal; Join MIS; and exit               /* round 2
      Listen for 1 round                                                      /* round 3
      **If** heard a signal in any round of exchange 2 **then** become *inactive*; and exit
    **end**
  **end**

Figure 2: Algorithm 1: MIS in the beeping model with collision detection.

---

**Lemma 1** *All messages received by node $j$ in the first exchange of step $i$ were sent by processors using the same probability as $j$ in that step.*

**Proof.** Let $k$ be a neighbor of $j$. If $k$ started at the same round as $j$ (both woke up at the same round) then they are fully synchronized and we are done. If $k$ started before $j$ then the first message $k$ sent has awakened $j$. Thus, they are only one round apart in terms of execution. Any message sent by $k$ in the 2nd round of the 1$st$ exchange of step $i$ would be received by $j$ in the first round of that exchange. Similarly, if $k$ was awakened after $j$ it must have been a 1 round difference and $j$ would receive $k$'s message for the first exchange of step $i$ (if $k$ decided to broadcast) in the third round of that exchange. Thus, all messages received by $j$ are from processors that are also in step $i$ and so all processors from which $j$ receives messages in that exchange are using the same probability distribution. ∎

Note that a similar argument would show that all messages received in the second exchange of step $i$ are from processors that are in the second exchange of that step. Since our safety proof only relies on the coherence of the exchange they still hold for this algorithm. Notice also that by adding a listening round at the beginning and end of each exchange the algorithm now works in the un-slotted model (with at most doubling the round complexity).

### 4.1 Message complexity

We conclude this section by presenting a proof that the total number of beep broadcast events in the algorithm is linear in the expected size of the resulting MIS set which is optimal since every node in the MIS must send at least one message as we show below. This is unlike in [1], where we counted the number of 'receive' events, i.e., each beep broadcast to $k$ active neighbors counts as $k$ in the message complexity. In [1] (lemma 4) we show that each 'broadcast' event has a constant probability (at least $\frac{1}{e^2}$) of resulting in the inclusion of the sender in the MIS set. Thus, if $m$ is the size of the MIS, the expected number of send events is $(me^2)$. Next, we prove that each MIS set member must send at least one message.

**Lemma 2** *Every node that enters the MIS broadcasts at least one message*

**Proof.** Assume there is an execution with $n$ nodes in which all nodes receive as input the upper bound $(n + 1)$ and in which a node $v$ enters the MIS set without broadcasting any message (if no such execution exists we are done). Let the round that $v$ enters the MIS in that execution be $r$. We now simulate another execution in which we add 1 node to the original set of nodes, call it $y$. We connect $y$ to all neighbors of $v$ and to $v$ itself. Since $v$ and $y$ are symmetric and since both $v$ and $y$ see the same set of messages up to and including round $r$ ($v$ does not broadcast any message so all messages both see are from $v$'s neighbors), there is a non negative probability that $y$ enters the MIS in round $r$ as well. However, since $v$ and $y$ are connected this violates the MIS guarantees. Thus, every node entering the MIS must broadcast at least one message. ∎

## 5 Beeping model

In the previous section we assumed a collision detection model that allowed nodes to listen to incoming messages in the same round they are sending. As noted in Section 3, the beeping model [5] as well as other communication models [12] are more strict and only allow listening in rounds in which a processor does not broadcast. To extend our algorithm to this model we replace the second round in exchange 1 of Algorithm 1 (Fig. 2) with $1 + c \log n$ rounds. In the first round of this new set each active processor flips a coin with the probability specified in Algorithm 1. If the flip outcome is 0 (tail) the processor does nothing for the next $c \log n$ rounds. If the flip outcome is 1 the processor sets $v = 1$ and selects at random (with equal probability) half of the rounds in the next $c \log n$ rounds. During the selected rounds the processor broadcasts a beep. In all other rounds it listens. If at any of the rounds it listens it hears a beep it sets $v = 0$ and stops broadcasting (even in the selected rounds). We say that two nodes are colliding in a specific round if they are neighbors and both have $v = 1$ at that round. Below we prove that if two or more nodes collide in the first round of exchange 1 (right after the coin flip) then with high probability they would not be colliding on the first round of exchange 2 of that step and so at most one of them enters the MIS. Notice however that there is some small probability ($\leq 1/n^{c/2-2}$) that two neighboring nodes enter the MIS simultaneously without hearing each other. Thus, unlike the other algorithms presented in this paper, in this variant the correctness of the MIS is probabilistic as well.

**Lemma 3** *Assume processor $y$ collided with one or more of its neighbors in the first round of exchange 1 in step $i$. Then the probability that $y$ would still be colliding with any of its neighbors in the first round of exchange 2 of that step is $\leq \frac{1}{n^{c/2}}$.*

**Proof.** If at any round in exchange 1 all neighbors of $y$ have v=0 we are done. Otherwise in each round that $y$ does not broadcast there is a probability of at least 0.5 that one of its colliding neighbors would broadcast leading $y$ to set $v = 0$. Since there are $(c \log n)/2$ rounds in which $y$ is listening, the probability that it would not hear a broadcast in any of these rounds is: $\leq \frac{1}{2^{c \log n/2}} = \frac{1}{n^{c/2}}$. ∎

Since there are only $O(\log^2 n) < n$ steps in our algorithm, and $n$ nodes, the probability that there exists a step and a node that collided during this step with a neighbor without resolving this collision is smaller than $\frac{1}{n^{c/2-2}}$. Thus, with probability $\geq 1 - \frac{1}{n^{c/2-2}}$ all collisions are detected and the MIS safety condition holds.

# 6 Reducing the dependency on the knowledge of the number of nodes

In this section we denote by $n$ the actual number of nodes in the graph and by $N$ the upper bound on the number of nodes provided to the algorithm. So far we assumed that $N$ is a polynomial factor of $n$ (which is why we used only $n$). Consider, however, cases in which $N \gg n$. This can occur in cases where the correct number is not known and also in cases where not all nodes participate (either to conserve energy or because they have other constrains). In such cases $N$ can be exponential in $n$. For example if $n \approx \log N$ the run time of Algorithm 1 would be $O(n^2)$. For such cases we present a variant of Algorithm 1 which runs in $O(\log n \log N)$ rounds. There is, of course, a price to be paid. In our case the price comes from an increase in message complexity as we discuss below. The algorithm is presented in Figure 3.

The main difference between this algorithm and Algorithm 1 is the inner loop which is performed in Algorithm 1 and omitted in this algorithm. Instead, each phase (the outer loop) performs $\log N$ steps and processors repeat the same set of steps until they exit the algorithm (either by entering the MIS or because they are connected to a node that enters the MIS). Since the exchanges are the same as in Algorithm 1 we use the collision detection model from Section 4 here for simplicity. As discussed in the previous section, with a factor of $O(\log N)$ we can extend this algorithm to the beeping model.

The safety property of our algorithm relies on the integrity of the exchanges and these are the same as in the Algorithm 1 (specifically, Lemma 1 holds for this algorithm as well). However, the run time is different as we show. We next prove that with high probability all nodes terminate the algorithm in $O(\log n \log N)$ time. Let $d_v$ be the number of active neighbors of node $v$. We start with the following definition [2]:

Definition: $v$ is Good if it has at least $d_v/3$ active neighbors $u$, s.t., $d_u \leq d_v$.

/* N is a given upper bound on the actual number of nodes in the network */

  Wait to hear the first wake up beep
  **If** received a beep (the first)
     **then** broadcast wake up beep to all neighbors    /* Make sure your neighbors wake up too
  Wait one round                          /* while your neighbor(s) wake up their neighbors
  **Repeat**                                 /*Until exiting the algorithm
    **for** $i = 0$ to $\log N$ **do:**                         /* $\log N$ phases
        **\*\* exchange 1 \*\* with** 3 **rounds**
      Listen for 1 round                                           /* round 1
      $v = 0$
      With probability $\frac{1}{2^{\log N - i}}$ broadcast signal and set $v = 1$ ;        /* round 2
      Listen for 1 round                                           /* round 3
      **If** heard a signal in any round of exchange 1 **then** $v = 0$
        **\*\* exchange 2 \*\* with** 3 **rounds**
      Listen for 1 round                                           /* round 1
      **If** $v == 1$ **then** Broadcast signal; Join MIS; and exit            /* round 2
      Listen for 1 round                                           /* round 3
      **If** heard a signal in any round of exchange 2 **then** become *inactive*; and exit
    **end**
  **end**

Figure 3: Algorithm 2: MIS in the beeping model *without* collision detection.

**Lemma 4.4 from [2]**: In every graph $G$ at least half of the edges touch a Good vertex. Thus, $\sum_{v \in Good} d_v \geq |E|/2$.

Below we show that $\log n$ phases are enough to guarantee that all processors terminate the algorithms with high probability.

**Lemma 4** *The expected number of edges deleted in a phase (with $\log N$ steps) is $\Omega(|E|)$*

**Proof.** Fix a phase $j$, and fix a Good vertex $v$. Assume that at the beginning of phase $j$, $2^k \leq d_v \leq 2^{k+1}$ for some $k < n$. If when we reach step $i = \log N - k$ in phase $j$ at least $d_v/20$ edges incident with $v$ were removed already we are done. Otherwise, at step $i$ there are still at least $d_v/3 - d_v/20 > d_v/4 \geq 2^{k-2}$ neighbors $u$ of $v$ with $d_u \leq d_v$. Node $v$ and all its neighbors $u$ are flipping coins with probability $\frac{1}{2^k}$ at this step and thus the probability that at least one of them would broadcast is:

$$p(v \ or \ u, \ neighbor \ of \ v, \ broadcasts) \geq 1 - (1 - \frac{1}{2^k})^{2^{k-2}} \cong 1 - 1/e^{1/4}$$

On the other hand, all nodes $u$, and $v$, have less than $2^{k+1}$ neighbors. Thus, the probability that a node from this set that broadcasts a message does not collide with any other node is:

$$p(no \ collisions) \geq (1 - \frac{1}{2^k})^{2^{k+1}} \cong 1/e^2$$

Thus, in every phase a Good node $v$ has probability of at least $(1 - \frac{1}{e^{1/4}})\frac{1}{e^2} \geq \frac{1}{2^7}$ to be removed. Thus, the probability that $v$ is removed is $\Omega(1)$ which means that the expected number of edges incident with $v$ removed during this phase is $\Omega(d_v)$.

Since half the edges touch a Good node, by the linearity of expectation the expected number of edges removed in each phase is $\geq \frac{1}{2}\Omega(\sum_{v \in Good} d_v) = \Omega(|E|)$.

Note that since the number of edges removed in a phase in a graph $(V, E)$ is clearly always at most $|E|$, the last lemma implies that for any given history, with probability at least $\Omega(1)$, the number of edges removed in a phase is at least a constant fraction of the number of edges that have not been deleted yet. Therefore there are two positive constants $p$ and $c$, both bounded away from zero, so that the probability that in a phase at least a fraction $c$ of the number of remaining edges are deleted is at least $p$. Call a phase successful if at least a fraction $c$ of the remaining edges are deleted during the phase.

By the above reasoning, the probability of having at least $k$ successful phases among $m$ phases is at least the probability that a binomial random variable with parameters $m$ and $p$ is at least $k$. By the standard estimates for Binomial distributions, and by the obvious fact that $O(\log |E|/c) = O(\log n)$ successful phases suffice to finish the algorithm, we conclude that with high probability the total running time of the algorithm is $O(\log n \log N)$.

∎

## 6.1 message complexity

Unlike Algorithm 1, the message complexity of this algorithm is not optimal. In fact, a lower bound for the message complexity is $O(n)$ even if the size of the MIS is $\Omega(1)$. To see this consider the complete graph with $n = N$ processors. Since only 1 processor can be in the MIS for this graph we need a step in which one processor flips a 1 while the rest obtain a 0. Thus, at each step $j$ ($j = 0, \ldots, \log N$) in phase 1 of algorithm 2 we have a probability of $n\frac{2^j}{n}(1 - \frac{2^j}{n})^{n-1}$ that the algorithm terminates. The probability that the algorithm would terminate in any of the steps in phase 1 is:

$$\leq \sum_{j=0}^{\log N} n\frac{2^j}{n}(1 - \frac{2^j}{n})^{n-1} \cong \sum_{j=0}^{\log N} \frac{2^j}{e^{2^j}} \leq \frac{1}{e} + \sum_{j=0}^{\log N} \frac{2}{e^2}e^{-2j} \leq 0.7$$

Thus, with a constant probability all nodes would reach the end of phase 1. At that step all nodes send messages with probability 1 and so the expected number of send events is at least $n$.

Note that this is tight, up to a logarithmic factor, for any graph on $n$ vertices. Indeed, the total number of phases is $O(\log N)$ and the expected number of broadcast messages a node sends during a phase is bounded by $\sum_i \frac{1}{2^i} < 2$, providing an $O(n \log N)$ upper bound.

# 7 A Round Complexity Lower bound

The algorithms we have presented so far shared a common theme: At each exchange all processors broadcast with the same probability. Note that this is not a requirement of the model and nodes can change their coin flip probability based on the history of messages they received which may lead to different probabilities for different nodes. However, for the lower bound proof we focus on such algorithms. Specifically, we assume synchronous rounds where at each exchange all processors flip coins with the same probability $p_t$ (though $p_t$ can be controlled by an adversary and set just before the exchange begins). We allow a round of coin flips that can be followed by several rounds of deterministic behavior and, combined, they consist of a single exchange in the algorithm. Our lower bound proof is based on the following graph construction. We consider a graph consisting of a disjoint union of complete bipartite graphs. For each $i$ between 1 and $(\log n)/4$ the graph has at least $n^{0.7}$ pairwise disjoint complete bipartite graphs of type $i$, that is, $n^{0.7}$ bipartite graphs with $2^i$ vertices in each side.

We first define the notion of a failed broadcast for a complete bipartite graph. We say that a broadcast failure has occurred in round $t$ for such graph if

1. No vertex in the graph broadcasts, or,

2. On both sides of the graph at least one vertex broadcasts a message.

**Lemma 5** *Let $G$ be a complete bipartite graph. If all rounds up to $t$ resulted in broadcast failure in $G$ then no vertex in $G$ has joined the MIS.*

**Proof.** We show that if a failure occurs up to and including round $t$ than there is a non zero probability that any node $j$ in $G$ has a completely symmetric neighbor $k$ in $G$ at round $t$. Thus, if $j$ joins the MIS there is a non zero probability that $k$ joins the MIS as well, in the same round, which violates the requirements for a MIS. We prove this by induction on $t$. *Base:* For $t = 0$ this is trivial and stems from our initial assumptions. For $t = 1$, if the failure is of type 1 all processors are clearly symmetric (both did not send or receive any message from the time the algorithm started). If the failure is of type 2 we distinguish between two cases. If $j$ has broadcasted in phase 1 let $k$ be a neighbor of $j$ that has also broadcasted in that round. Both $j$ and $k$ send and receive a message and so they are symmetric. If $j$ did not broadcast at phase 1 than $p_t < 1$ and there is a positive probability that there exists a node $k$ that is a neighbor of $j$ and did not broadcast at phase 1.

*Induction step:* We assume correctness for $t - 1$ and prove for $t$. If the failure at round $t$ is of type 1 then all nodes that were symmetric in phase $t - 1$ remain symmetric in phase $t$. If the failure is of type 2 then let $j$ and $k$ be two symmetric neighbors at the beginning of phase $t$. Since both received messages and both use the same probability for broadcasting an argument similar to the one used above shows that there is a positive probability that they would remain symmetric at the end of this phase as well. ∎

We now compute the probability that broadcast failure occurs for a specific type of a bipartite graph. Let $p_t^i$ be the probability of such broadcast failure for one complete bipartite graph $G$ with $2^i$ vertices on each side at phase t. Assume the broadcast probability for this phase is $p = p(t)$. Then:

$$p_t^i = (1-p)^{2^{i+1}} + \left(1 - (1-p)^{2^i}\right)^2$$

The left hand side of this sum accounts for failure of type 1 (no node broadcasts) while the right hand side accounts for failures of type 2.

If $p = 0$ then we are guaranteed that a broadcast failure of type 1 occurs in $G$. Otherwise let $k$ be an integer such that $\frac{1}{2^{k+1}} < p \leq \frac{1}{2^k}$. For $j = 0, \ldots, (k-1)$ we use the left side of the above sum to obtain a lower bound on the broadcast failure probability:

$$(1-p)^{2^{(j+1)}} \geq \left(1 - \frac{1}{2^k}\right)^{2^{j+1}} \cong \left(\frac{1}{e}\right)^{\frac{1}{2^{k-j-1}}}$$

For $j = k+1, \ldots, (\log n)/4$ we use the right hand side of the sum to bound the broadcast failure probability:

$$\left(1 - (1-p)^{2^j}\right)^2 = 1 - 2(1-p)^{2^j} + (1-p)^{2^{j+1}} \geq 1 - 2(1-p)^{\frac{p}{p}2^j} = 1 - 2\left((1-p)^{1/p}\right)^{p2^j}$$

$$= 1 - 2e^{-p2^j} \geq 1 - 2e^{-2^j/2^k} = 1 - 2e^{-2^{j-k}}$$

Finally, for $j = k$ we use the left hand side of the sum again to get:

$$(1-p)^{2^{j+1}} \geq \left(1 - \frac{1}{2^k}\right)^{2^{k+1}} \cong \frac{1}{e^2} > 0.1$$

We now compute the product of $p_t^j$ over all possible values of $j$ ($0 \leq j \leq \log n/4$) for a fixed phase $t$ (again we set $p = p(t)$). This product is:

$$\prod_{j=0}^{\log n/4} p_t^j \geq \prod_{j=0}^{\log n/4} \left((1-p)^{2^{(j+1)}} + \left(1 - (1-p)^{2^j}\right)^2\right)$$

$$\geq \left(\prod_{j=0}^{k-1} (1/e)^{\frac{1}{2^{k-j-1}}}\right) 0.1 \left(\prod_{j=k+1}^{(\log n)/4} 1 - 2e^{-2^{j-k}}\right)$$

The left part of this product evaluates to:

$$\left(\prod_{j=0}^{k-1} (1/e)^{\frac{1}{2^{k-j-1}}}\right) = \left(\frac{1}{e}\left(\frac{1}{e}\right)^{1/2}\left(\frac{1}{e}\right)^{1/4}\ldots\left(\frac{1}{e}\right)^{\frac{1}{2^{k-1}}}\right) \geq \frac{1}{e^2}$$

The right part is:

$$\left(\prod_{j=k+1}^{(\log n)/4} 1 - 2e^{-2^{j-k}}\right) \geq 1 - 2\left(\prod_{j=k+1}^{(\log n)/4} e^{-2^{j-k}}\right) \geq 1 - 2\left(\prod_{j=k+1}^{(\log n)/4} e^{-2(j-k)}\right)$$

$$\geq 1 - 2\left(\frac{1}{e^2(1-e^{-2})}\right) > 0.1$$

The first inequality stems from the fact that $(1-a)(1-b) \geq (1-a-b)$. Thus, for the full product we have:

$$\prod_{j=0}^{\log n/4} p_t^j \geq \prod_{j=0}^{\log n/4} \left((1-p)^{2^{(j+1)}} + \left(1 - (1-p)^{2^j}\right)^2\right) \geq \frac{1}{e^2} * 0.1 * 0.1 \geq .001 \geq \frac{1}{2^{10}}$$

Thus, if we have $T$ phases, no matter what $p_t$ is in each phase, the product over $p_t^j$ for all values of $j$ and all phases $1 \leq t \leq T$ is:

$$\prod_{t=1}^{T} \prod_{j=0}^{(\log n)/4} p_t^j \geq \frac{1}{2^{10T}}$$

By changing order in this product we get:

$$\prod_{t=1}^{T} \prod_{j=0}^{(\log n)/4} p_t^j = \prod_{j=0}^{(\log n)/4} \prod_{t=1}^{T} p_t^j = \prod_{j=0}^{(\log n)/4} \left(\prod_{t=1}^{T} p_t^j\right) \geq \frac{1}{2^{10T}} = \prod_{j=0}^{(\log n)/4} \left(\frac{1}{2^{10T}}\right)^{4/(\log n)}$$

Thus, for at least one value of $j$ we have

$$\prod_{t=1}^{T} p_t^j \geq \left(\frac{1}{2^{10T}}\right)^{4/(\log n)}$$

Set $T = 0.01 \log^2 n$. For such $j$ and $T$ we have:

$$\prod_{t=1}^{T} p_t^j \geq \left(\frac{1}{2^{10T}}\right)^{4/(\log n)} \geq \frac{1}{\sqrt{n}}$$

The above inequality shows that there exists a $j$ such that the probability that a single bipartite graph of type $j$ experiences a broadcast failure in all rounds is at least $\frac{1}{\sqrt{n}}$. Hence the probability it does not experience such failure is at most $(1 - \frac{1}{\sqrt{n}})$. As there are at least $n^{0.7}$ graphs of this type, the probability that none of them fails in all rounds is at most $(1 - \frac{1}{\sqrt{n}})^{n^{0.7}} \cong (1/e)^{n^{0.2}}$. This shows that almost surely there is a bipartite graph that will experience a broadcast failure in all $T$ phases.

Thus we need $\Omega(\log^2 n)$ phases for any algorithm solving the MIS problem under this model.

**Comment:** In the discussion above we assumed that only one probability value is used by all nodes in each coin flipping round. However, this assumption can be relaxed to allow $c$ different probabilities to be used in each round (for some constant $c$). For such a model we allow processors to select among a predefined set of $c$ probabilities for *each* phase (they can be different between the phases) based on the history of the messages they received. Note that all nodes in a complete bipartite graph that experienced a broadcast failure up to phase $t$ will use the same probability at that phase since all nodes in such graphs see the same set of incoming messages up to phase $t$. To show that our proof still holds note that replicating each graph $c$ times when computing the broadcast failure probability leads to a failure probability of $\frac{1}{2^{10c}}$ for each set. We then use $T = (1/100c)(\log^2 n)$ for the reminder of the proof.

## 8   Conclusions and future work

We have shown that using a severely constrained communication model based on a biological process, we can derive efficient algorithms for the MIS selection problem. Our algorithm can work with two different types of collision detection models. We have also presented a lower bound for a variant of the fly model in which coin flipping probability is constrained so that all processors flip coins with the same probability in each round. MIS Algorithm 1 was inspired by insights from the SOP selection process in flies, which solves a variant of the MIS selection problem [25]. Other biological systems, including various types of networks in the cell, are also addressing distributed computing problems. These include resource allocation during cell division [19], routing [26] and backup and fault tolerance in regulatory networks [8]. It is our hope that new ideas for improving other distributed computing tasks can be derived from studies of these biological processes.